\documentclass[12pt]{iopart}
\usepackage{graphicx}
\begin{document}

\title{Trapping hot quasi-particles in a high-power superconducting electronic cooler}

\author{H. Q. Nguyen}
\address{Low Temperature Laboratory (OVLL), Aalto University School of Science, P.O. Box 13500, 00076 Aalto, Finland}
\ead{hung@ltl.tkk.fi}
\author{T. Aref}
\address{Low Temperature Laboratory (OVLL), Aalto University School of Science, P.O. Box 13500, 00076 Aalto, Finland}
\ead{thomasaref@gmail.com}
\author{V. J. Kauppila}
\address{Low Temperature Laboratory (OVLL), Aalto University School of Science, P.O. Box 13500, 00076 Aalto, Finland}
\ead{ville.kauppila@aalto.fi}
\author{M. Meschke}
\address{Low Temperature Laboratory (OVLL), Aalto University School of Science, P.O. Box 13500, 00076 Aalto, Finland}
\ead{meschke@boojum.hut.fi}
\author{C. B. Winkelmann}
\address{Institut N\'eel, CNRS, Universit\'e Joseph Fourier and Grenoble INP, 25 avenue des Martyrs, 38042 Grenoble, France}
\ead{clemens.winkelmann@grenoble.cnrs.fr}
\author{H. Courtois}
\address{Institut N\'eel, CNRS, Universit\'e Joseph Fourier and Grenoble INP, 25 avenue des Martyrs, 38042 Grenoble, France}
\ead{herve.courtois@grenoble.cnrs.fr}
\author{J. P. Pekola}
\address{Low Temperature Laboratory (OVLL), Aalto University School of Science, P.O. Box 13500, 00076 Aalto, Finland}
\ead{jukka.pekola@aalto.fi}
\begin{abstract}
The performance of hybrid superconducting electronic coolers is usually limited by the accumulation of hot quasi-particles in their superconducting leads. This issue is all the more stringent in large-scale and high-power devices, as required by applications. Introducing a metallic drain connected to the superconducting electrodes via a fine-tuned tunnel barrier, we efficiently remove quasi-particles and obtain electronic cooling from 300 mK down to 130 mK with a 400 pW cooling power. A simple thermal model accounts for the experimental observations.
\end{abstract}
\maketitle

\section*{Introduction}
On-chip solid-state refrigeration has long been sought for various applications in the sub-kelvin temperature regime, such as cooling astronomical detectors \cite{Moseley,Richards,MillerAPL08}. In a Normal metal - Insulator - Superconductor (NIS) junction \cite{NahumAPL,MuhonenRPP,GiazottoRPM06}, the superconductor density of states gap makes that only high-energy electrons are allowed to tunnel out of the normal metal or, depending on the bias, low-energy ones to tunnel in, so that the electronic bath as a whole is cooled. In SINIS devices based on aluminum, the electronic temperature can drop from 300 mK down to below 100 mK at the optimum bias point. While this level of performance has been demonstrated in micron-scale devices \cite{PekolaPRL04,RajauriaPRL07} with a cooling power in the picoWatt range, a difficulty arises in devices with large-area junctions needed for a sizable cooling power approaching the nanoWatt range. For instance, a high-power refrigerator has been shown to cool an external object from 290 mK down to about 250 mK \cite{UllomAPL13}. One of the main limitation to NIS coolers' full performance is the presence in the superconducting leads of non-equilibrium quasi-particles arising from the high current running through the device. The low quasi-particle relaxation rate and thermal conductivity in a superconductor bound these hot particles in the vicinity of the junction and lead to severe overheating in the superconducting electrodes.

There are several methods for reducing the accumulation of quasi-particles in a superconductor. For example a small magnetic field \cite{PeltonenPRB10} can be used to introduce vortices that trap quasi-particles. This approach is however not applicable to electronic coolers with large-area junctions since a vortex also reduces the cooling performance if it resides within a junction. The most common method is to use a normal metal coupled to the superconductor as a quasi-particle trap: quasi-particles migrate to the normal metal and relax their energy there through electron-electron and electron-phonon interaction. In the typical case of a fabrication process based on angle evaporation, quasi-particle traps are formed by the structures mirroring each superconducting electrode, sitting on a side of the cooling junction and featuring the same oxide barrier layer. The trapping efficiency is usually moderate, but can be improved in two ways: the normal metal can be put in direct contact with the superconductor, as out-of-equilibrium quasi-particles would diffuse more efficiently to the trap \cite{Agulo04}, or the trap can be closer to the junction \cite{ONeilPRB12,Luukanen}. In both cases, it is important to prevent inverse proximity effect in the superconductor, which smears locally the superconductor density of states and degrades cooling efficiency. The existence of an optimum transparency for the interface between the trap and the superconducting lead is therefore expected, but remains to be investigated \cite{Kauppila}.

In this paper, we present an effective method to evacuate quasi-particles in a SINIS cooler, based on what we call a quasi-particle drain. It is a kind of quasi-particle trap made of a layer of normal metal located at a fraction of superconducting coherence length away from the junction and separated by a thin insulating layer to stop the inverse proximity effect. We compare the cooling performance when varying the quasi-particle drain barrier transparency over a wide range. The efficiency of the quasi-particle drain is demonstrated through the electronic cooling from 300 to 130 mK at a 400 pW cooling power. A simple thermal model captures the effect of the quasi-particle drain reasonably well.

\section*{Fabrication and measurement methods}

\begin{figure}[tb]
\begin{center}
\includegraphics[width=0.8\columnwidth,keepaspectratio]{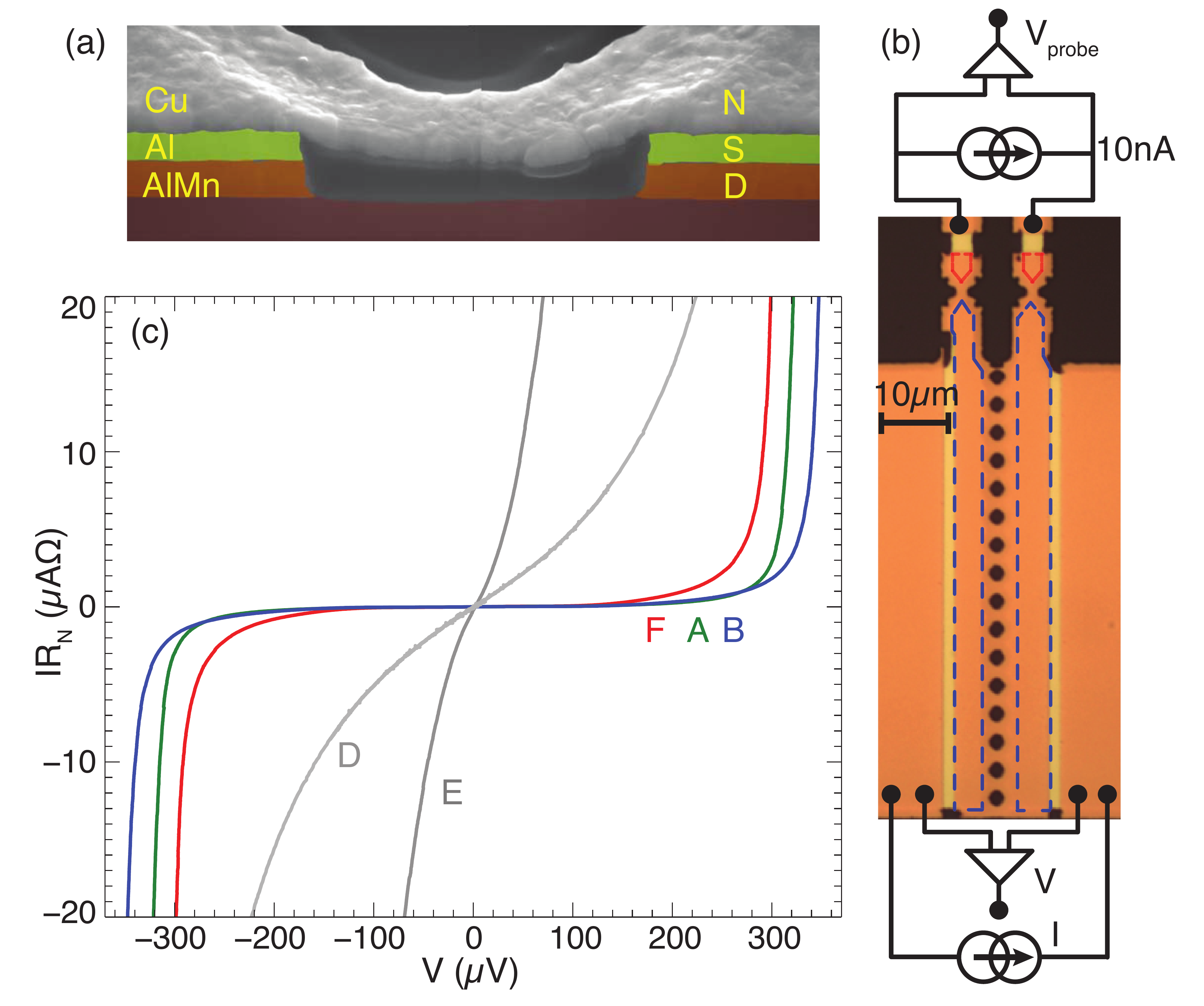}
\caption{(a) False-colored scanning electron side micrograph of a SINIS cooler showing from top to bottom: 100 nm of Cu (cooled normal metal N), 200 nm of Al (superconductor S), and 200 nm of AlMn (quasi-particle drain D) with AlOx insulating layers in-between (not visible). The Cu layer is suspended on the top of the Al/AlMn layers. (b) Top view of the cooler with the measurement setup. The array of holes connects the two NIS junctions, whose size is 70$\times$4 $\mu$m$^2$. Cu appears as orange and the Al/AlMn bilayer as yellow. The cooler junctions are highlighted by blue dashed lines, and the thermometer junctions are highlighted by red dashed lines. The circuit connected to the top electrodes probes electron temperature of the normal island cooled by the two current-biased large junctions. (c) Current-voltage characteristics at a 50 mK temperature of samples A, B, D, E and F with different tunnel barrier thicknesses between the quasi-particle drain and the superconducting leads, see Table 1.}
\label{fig1}
\end{center}
\end{figure}

We use the fabrication process described in \mbox{Ref.} \cite{NguyenAPL}, in which a SINIS cooler is obtained by photo-lithography and chemical etch of a NIS multilayer. Here, we add a normal metal layer at the bottom, which is used as a quasi-particle drain. Figure 1a false-colored scanning electron micrograph shows a side view of a typical SINIS cooler, obtained by cutting it with a focused ion beam. From top to bottom, the 100 nm-thick Cu normal metal layer to be cooled is suspended between two 200 nm-thick Al superconducting electrodes. The latter rest on two separate quasi-particle drains made of a 200 nm layer of AlMn deposited on a Si wafer. We choose AlMn \cite{AlMn,ClarkAPL} as a quasi-particle drain normal material because it acts chemically as Al in terms of oxidation and etch. The layers are separated by two aluminum oxide barriers, which we name the drain barrier between the AlMn and Al layers and the cooler barrier between the Al and Cu layers. Sample parameters are given in Table 1.

\small
\begin{table}
\caption{\label{table}SINIS cooler parameters. All coolers are made of layers of 100 nm of Cu, 200 nm of Al, and 200 nm of AlMn and have a junction size of 70$\times$4 $\mu$m$^2$. As an exception, sample F has no AlMn layer as a quasi-particle drain and the thickness of Al is 400 nm. We indicate oxygen pressure and oxidation time used for the preparation of the AlMn/Al drain barrier and of the Al/Cu cooler barrier. 2$R_N$ and $2\Delta$ are the normal state resistance and twice the superconducting energy gap, respectively, obtained by fitting the IV characteristics to \mbox{Eq. 1} for SINIS structures. "Color" refers to the figures throughout the paper.}
\begin{indented}
\item[]\begin{tabular}{@{}lllllll}
\br
Sample & Drain barrier & Cooler barrier & $2R_N$ & $2\Delta$ &Color\\
 & (mbar, second) & (mbar, second) & $(\Omega)$ & $(\mu eV)$ & & \\
\mr
A & 1.3, 10 & 1.3, 300 & 0.71 & 398  & green \\%w45
B & 0.26, 10 & 1.3, 300 & 1.56 & 382 & blue \\%, w49
C & 0.18, 1 & 0.8, 180 & 0.55 & 370 & purple\\%, w51
D & 5$\times$10$^{-4}$, 10 & 1.3, 300 & 1.01 & 228 & gray \\%, w52
E & 0 & 1.3, 180 & 1.31 & 180 & gray \\%, w46
F & N/A & 1, 300 & 0.83 & 390 & red\\%, w23
\br
\end{tabular}
\end{indented}
\end{table}
\normalsize

Figure 1b is an optical micrograph showing a top view of the cooler. The two NIS junctions area are outlined by dashed blue lines. They are separated by a trench in the Al and AlMn layers, created by chemical over-etch, underneath the array of holes in the suspended Cu layer. Each junction has an area of 70 $\times$ 4 $\mu$m$^2$ and is surrounded by two quasi-particle traps: a side trap made of Cu next to it, and a quasi-particle drain made of AlMn. Two additional small NIS junctions connected to the normal metal are used as a SINIS thermometer. Electron temperature is accessed by comparing the measured voltage $V_{probe}$ drop under a small bias current (typically 10 nA) to a calibration against the cryostat temperature.

\section*{Experimental results and discussion}
The current flowing through a NIS junction with a voltage V writes
\begin{equation}
I_{NIS}=\frac{1}{eR_N}\int{dE n_S(E)[f_N(E-eV)-f_S(E)]},
\end{equation}
where $n_S=Re[E/\sqrt{E^2-\Delta^2}]$ is the normalized superconductor density of states, $\Delta$ is the superconducting gap, $R_N$ is the normal state junction resistance, and $f_{S,N}$ are the Fermi-Dirac energy distributions of electrons in S and N, respectively. If leakage is negligible, a low sub-gap current then means a low electronic temperature.

Figure 1c shows the current voltage characteristics (IV curves) of different samples measured with a standard current-biased 4-probe technique in a dilution cryostat at 50 mK, with a focus on the low-bias regime. The two innermost curves stand for samples E and D, which have no drain barrier and a very thin drain barrier, respectively. Superconductivity in the Al layers is then affected by a strong inverse proximity effect, which results in a depressed critical temperature and a low superconducting gap $\Delta$ so that $2\Delta$ = 180 $\mu$eV and 228 $\mu$eV respectively.

As samples A, B, and C are fabricated using a higher oxidation pressure for the drain barrier, they have a typical value for $2\Delta$ of about 350 $\mu$eV and a ratio of minimum conductance to normal-state conductance of about $10^{-4}$, indicating that inverse proximity effect is weak. These samples show a sharp IV characteristic, with sample C (not shown) behaving almost identically to sample B. The sole difference between samples A and B is the drain barrier. Having a thinner barrier, sample B exhibits less current at a given sub-gap bias, \mbox{i.e.} less overheating from quasi-particles in the superconductor. The drain barrier lets quasi-particles get efficiently trapped in the drain, while it stops the inverse proximity effect.

\begin{figure}[t]
\begin{center}
\includegraphics[width=0.9\columnwidth,keepaspectratio]{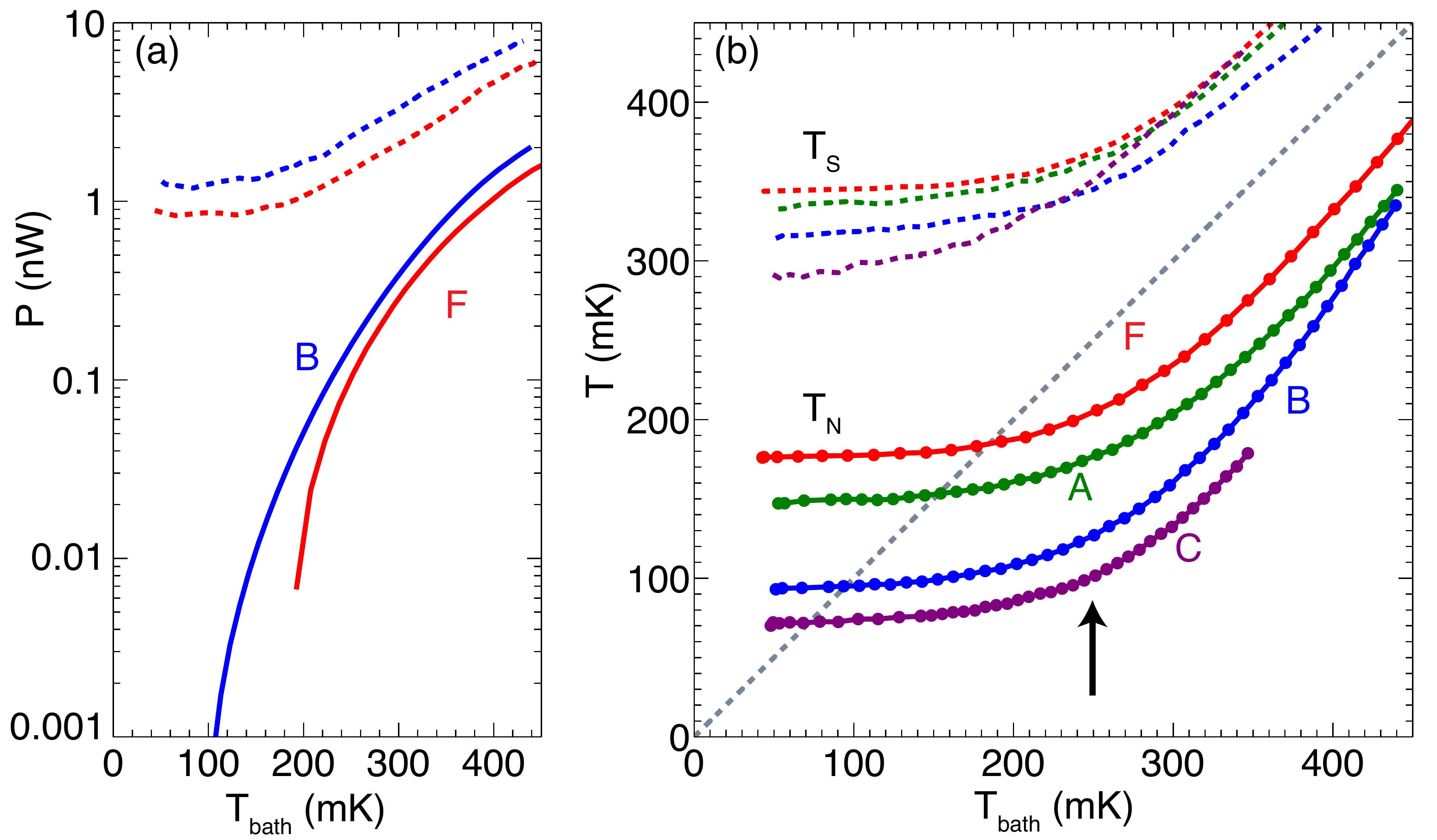}
\caption{(a) Cooling power (solid lines) and input power (dashed lines) of cooler B (blue) and F (red) as a function of the bath temperature. (b) Calculated electron temperature (dashed lines) of the superconductor and measured normal metal temperature (dots, connected by solid lines as a guide to the eye) at the optimum point for samples A, B, C and F. The dashed gray line is a one to one line, indicating the bath temperature. The arrow marks the bath temperature of 250mK, where sample C cools down to below 100 mK.}
\label{cooling}
\end{center}
\end{figure}

In a normal metal, the main heat flux from the electron system to the environment is through the coupling with the phonon system. We estimate the cooling power from the electron-phonon coupling power using
\begin{equation}
\dot{Q}_{ep}=\Sigma^N \mathcal V(T_N^5-T_{ph}^5),
\end{equation}
where $\Sigma^N=2\times 10^9$ WK$^{-5}$m$^{-3}$ is the electron-phonon coupling constant in Cu \cite{WellstoodPRB,MeschkeJLTP04}, $\mathcal V=83$ $\mu$m$^3$ is the island volume for all samples, $T_N$ is the measured electron temperature and $T_{ph}$ is the phonon temperature, both in the normal island. We assume here that $T_{ph}=T_{bath}$, which leads to an upper limit on the estimation of the cooling power. Figure 2a displays the calculated electron-phonon coupling power $\dot{Q}_{ep}$ (solid lines) and the total Joule power (dashed lines) $P_{IV}=I_{opt} V_{opt}$ measured at the optimum bias, \mbox{i.e.} the bias at which electronic cooling is maximum. The typical power scale for all our present coolers is in the nanoWatt range.

Figure 2 b presents the most important result of our work, namely the behavior of the normal island electronic temperature at optimum bias as a function of the bath temperature. All coolers have the largest temperature drop at a bath temperature around 300 mK. At lower temperatures, heating above the bath temperature is observed, which arises from the sub-gap current. Having only the side trap and no quasi-particle drain, sample F shows a poor cooling from 300 mK down to only 230 mK. Carrying a quasi-particle drain, sample A cools to 200 mK. With a reduced drain barrier, sample B cools to 160 mK. In sample C, we obtain the best cooling by further reducing the drain and the cooler barriers: from 300 mK, sample C cools down to 132 mK with a 400 pW cooling power. Most importantly, from a bath temperature of 250 mK, sample C cools down to below 100 mK, see the arrow in \mbox{Fig. 2b}. 

The electronic temperature of the superconducting electrodes can be accessed by balancing the normal metal electron-phonon coupling power (Eq. 2) with the NIS junction cooling power at a voltage bias $V$:
\begin{equation}
\dot{Q}_{NIS}=\frac{1}{e^2R_N}\int{dE(E-eV)n_S(E)[f_N(E-eV)-f_S(E)]}.
\end{equation}
Figure 2 b upper part (dashed lines) displays the electronic temperature of the superconductor $T_S$ derived when assuming again $T_{ph}=T_{bath}$ in the normal metal. Although the latter assumption calls for further discussion, the trend remains that the superconductor gets significantly overheated. 

\section*{Thermal model}

\begin{figure}[t]
\begin{center}
\includegraphics[width=0.9\columnwidth,keepaspectratio]{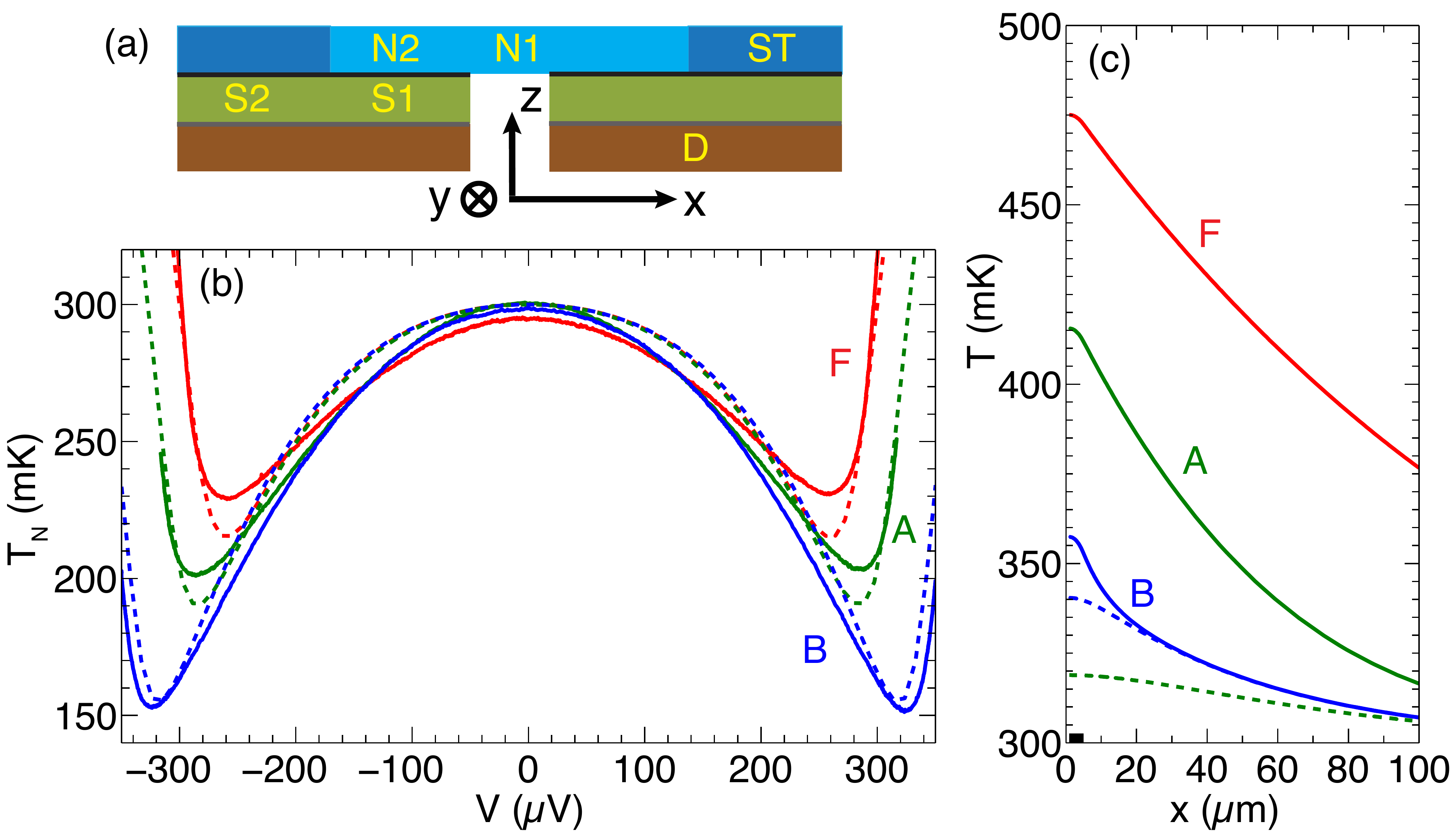}
\caption{Finite element study of heat transport in coolers A, B and F. (a) Geometry of the model: N1 is the suspended part of the normal island, N2 stands for the junction areas, S for the superconducting electrodes (S1 under the junctions and S2 under the side traps), ST for the side traps, and D for the quasi-particle drains. The layers are separated by tunnel barriers. (b) Electronic temperature drop as a function of bias at a 300 mK bath temperature, solid lines are measured data while dashed lines are calculated from the model. (c) Temperature of the superconductor S (solid lines) and of the quasi-particle drain D (dashed lines) as a function of the distance from the junction. The black bar indicates the junction location.}
\label{comsol}
\end{center}
\end{figure}

In order to describe further the thermal transport in our devices, we consider a one-dimensional multilayered thermal model. It is a set of coupled heat equations for different systems. The model geometry shown in figure 3a includes a normal metal island (N), superconducting leads (S), side traps (ST), and quasi-particle drains (D), similar to the geometry of the studied coolers (figure 1a). Although a non-equilibrium description of a biased NIS junction is possible \cite{VoutilainenPRB,SukumarPRB09}, we assume for simplicity every part of the device can be described by local temperature. We also neglect inverse proximity effect in the superconductor density of electronic states, as can be justified by the sharpness of measured IVs.

In the following equations, we will make use of the quantities $\mathcal I_{NIS}$, $\mathcal P_{NIS}$ and $\mathcal P^i_{ep}$ that are the local current $I_{NIS}$, cooling power $\dot{Q}_{NIS}$ and electron-phonon coupling power $\dot{Q}_{ep}$ per unit area in the $xy$ plane, see Fig. 3a. In addition, $T_i$ is the temperature, $d_i$ is the thickness and $\kappa_i$ is the thermal conductivity in the element $i$. Here, temperature gradients are in the $x$-direction, charge and heat currents through the junctions ($\mathcal I_{NIS}$ and $\mathcal P_{NIS}$) are in the $z$-direction, electron-phonon couplings $\dot{Q}_{ep}$ are in the bulk. In the normal metal, the temperature gradient is determined by the electronic thermal conductivity, the local electron-phonon coupling plus, in the region N2 in contact with the junction, the local cooling power as
\begin{eqnarray}
N1: \kappa_N d_N \nabla^2T_N&=&-\mathcal P_{ep}^N(T_N),\nonumber\\
N2: \kappa_N d_N \nabla^2T_N&=&\mathcal P_{NIS}(V,T_S,T_N,R_N)-\mathcal P_{ep}^N(T_N).
\end{eqnarray}
In the superconductor, the cooling junction injects a heat current	 equal to the Joule power plus the cooling power. The effect of the drain in region S1 or of the side trap in region S2 is described using the expression of the heat flow through a NIS junction at zero bias $\mathcal P_{NIS}(V=0)$. Weak electron-phonon coupling in S1 and S2 is neglected here, and
\begin{eqnarray}
S1: d_S \nabla.(\kappa_S\nabla T_S)&=&\mathcal I_{NIS}(V,T_S,T_N,R_N)V-\mathcal P_{NIS}(V,T_S,T_N,R_N)\nonumber\\
&&-\mathcal P_{NIS}(0,T_S,T_{D},R_{D}),\nonumber\\
S2: d_S \nabla.(\kappa_S\nabla T_S)&=&-\mathcal P_{NIS}(0,T_S,T_{D},R_{D})-\mathcal P_{NIS}(0,T_S,T_{ST},R_N).
\end{eqnarray}
The drain is submitted to the heat coming from the superconductor through the drain barrier. As it is made of a normal metal, electron-phonon coupling is taken into account as
\begin{eqnarray}
\kappa_D d_D \nabla^2T_{D}&=&\mathcal P_{NIS}(0,T_S,T_{D},R_{D})-\mathcal P_{ep}^D(T_D),
\end{eqnarray}
and similarly for the side trap
\begin{eqnarray}
\kappa_N d_N \nabla^2T_{ST}&=&\mathcal P_{NIS}(0,T_S,T_{ST},R_N)-\mathcal P_{ep}^{ST}(T_{ST}).
\end{eqnarray}

We solved these coupled differential equations numerically \cite{comsol} using measured parameters from samples A, B, and F at 300 mK. From the measured electrical conductivity, we find that our sputtered Cu films have a residual resistance ratio of about 1.5, which is limited by disorder in the film \cite{Qian}. The deduced value of $\kappa_0=0.9$ WK$^{-2}$m$^{-1}$ ($\kappa_N=\kappa_0T$) is in agreement with the tabulated value in \cite{GiazottoRPM06}. We use $\kappa_D=0.47 \kappa_N$ \cite{AlMn} and $\Sigma^D=10^9$ WK$^{-5}$m$^{-3}$ \cite{ClarkAPL}. We take into account the exponential decay of $\kappa_S$ with temperature \cite{Timofeev}. For the cooler barrier, we use $R_N$ = 500 $\Omega \mu$m$^2$, close to the measured value for sample A (400 $\Omega \mu$m$^2$) and B (700 $\Omega \mu$m$^2$). Based on the different properties of samples A and B (Table 1), we use drain barrier resistivity $R_D$ = 500 $\Omega \mu$m$^2$ for sample A and $R_D$ = 10 $\Omega \mu$m$^2$ for sample B. Note that $R_D$ is the only differing parameter between samples A and B. This guess (based on the value of the cooler barrier and the prediction in \cite{Kauppila}) is necessary as we cannot measure $R_D$ directly. In solving for sample F, equation (6) and the terms for $T_D$ in equation (5) are ignored as F does not have a quasi-particle drain.

Solving these equations gives a complete temperature profile: $T_N$, $T_S$, $T_D$, and $T_{ST}$ of the device. Figure 3b compares $T_N$ as a function of bias voltage from the modeling results (dashed lines) with the experimental data (solid lines) at 300 mK. The good match between the two confirms that our simple model captures the essential physics of the device: a thinner drain barrier is the single parameter that enhances the performance of the cooler. Figure 3c shows the calculated temperature profile in the superconductor (solid lines) and in the drain (dashed lines) at optimum bias. It consistently shows that superconducting electrodes get overheated over a typical length scale of about 50 $\mu$m from the junction. Carrying a weak drain barrier transparency, sample A has a local superconductor temperature $T_S$ well above the drain temperature $T_D$. With an improved barrier transparency in sample B, its superconducting electrodes are well thermalized by the drains. One obtains $T_S\approx T_D$ at distances x of about 20 $\mu$m away from the junction. The behavior is consistent with the magnitude of electronic cooling observed, thus demonstrating the effect of an improved drain barrier transparency for a good efficiency of the quasi-particle drain.

\section*{Conclusions}
We have designed and studied electronic coolers capable to cool an electronic bath from 250 mK, the base temperature of a He$^3$ cryostat, down to below 100 mK, the working regime of a dilution cryostat. With a fine-tuned barrier, the quasiparticle drain is efficient in thermalizing NIS junctions. The related geometry does not impose any limit in making the junction larger, thus opening the possibility to obtain a cooling power well above the present level of 400 pW. The fabrication is low cost and involves only photolithography, the device is of high quality and robust. On this basis, we are developing a platform that integrates coolers and sensors on a single chip, which is of great potential for astrophysics and other low temperature applications. 

\section*{Acknowledgments}
We acknowledge the support of the European Community Research Infrastructures under the FP7 Capacities Specific Programme, MICROKELVIN project number 228464, the EPSRC grant EP/F040784/1, and the Academy of Finland through its LTQ CoE grant (project no. 250280). Samples are fabricated in the Micronova Nanofabrication Center of Aalto University. We thank D. Gunnarsson for help with the sputter.


\section*{References}
\begin{thebibliography}{30}
\bibitem{Moseley} S. H. Moseley, AIP Conf. Proc. \textbf{1185}, 9 (2009).
\bibitem{Richards} P. L. Richards, J. Appl. Phys. \textbf{76}, 1 (1994).
\bibitem{MillerAPL08} N. A. Miller, G. C. O'Neil, J. A. Beall, G. C. Hilton, K. D. Irwin, D. R. Schmidt, L. R. Vale, and J. N. Ullom, Appl. Phys. Lett. \textbf{92}, 163501 (2008).
\bibitem{NahumAPL} M. Nahum, T. M. Eiles and J. M. Martinis, Appl. Phys. Lett. \textbf{65}, 3123 (1994).
\bibitem{MuhonenRPP} J. T. Muhonen, M. Meschke, and J. P. Pekola, Rep. Prog. Phys. \textbf{75}, 046501 (2012).
\bibitem{GiazottoRPM06} F. Giazotto, T. T. Heikkila, A. Luukanen, A. M. Savin and J. P. Pekola, Rev. Mod. Phys. \textbf{78}, 217 (2006).
\bibitem{PekolaPRL04} J. P. Pekola, T. T. Heikkila, A. M. Savin, J. T. Flyktman, F. Giazotto, and F. W. J. Hekking, Phys. Rev. Lett. \textbf{92}, 056804 (2004).
\bibitem{RajauriaPRL07} S. Rajauria, P. S. Luo, T. Fournier, F. W. J. Hekking, H. Courtois, and B. Pannetier, Phys. Rev. Lett. \textbf{99}, 047004 (2007).
\bibitem{UllomAPL13} P. J. Lowell, G. C. O'Neil, J. M. Underwood, and J. N. Ullom, Appl. Phys. Lett. \textbf{102}, 082601 (2013)
\bibitem{PeltonenPRB10} J. T. Peltonen, J. T. Muhonen, M. Meschke, N. B. Kopnin, and J. P. Pekola, Phys. Rev. B \textbf{84}, 220502 (2010).
\bibitem{Agulo04} I. J. Agulo, L. Kuzmin, M. Fominsky, M. Tarasov, Nanotechnology \textbf{15}, S224 (2004).
\bibitem{ONeilPRB12} G. C. O'Neil, P. J. Lowell, J. M. Underwood, and J. N. Ullom, Phys. Rev. B \textbf{85}, 134504 (2012).
\bibitem{Luukanen} A. Luukanen, A. M. Savin, T. I. Suppula, J. P. Pekola, M. Prunnila, and J. Ahopelto, LTD-9 AIP Conf. Proc. \textbf{605}, 375-378 (2002).
\bibitem{Kauppila} V. J. Kauppila, H. Q. Nguyen and T. T. Heikkila,  arXiv:1304.1288, under review in Phys. Rev. B.
\bibitem{NguyenAPL} H. Q. Nguyen, L. M. A. Pascal, Z. H. Peng, O. Buisson, B. Gilles, C. B. Winkelmann, and H. Courtois, Appl. Phys. Lett. \textbf{100}, 252602 (2012).
\bibitem{AlMn} Al doped Mn at 2500 ppm. We have tested down to 50 mK that its resistivity $\rho=$ 0.036 $\Omega\mu$m (= $\rho_{Cu}/0.47$).
\bibitem{ClarkAPL} A. M. Clark, A. Williams, S. T. Ruggiero, M. L. van den Berg, and J. N. Ullom, Appl. Phys. Lett. \textbf{84}, 625 (2004).
\bibitem{WellstoodPRB} F. C. Wellstood, C. Urbina, and J. Clarke, Phys. Rev. B {\bf 49}, 5942 (1994).
\bibitem{MeschkeJLTP04} M. Meschke, J. P. Pekola, F. Gay, R. E. Rapp, and H. Godfrin, J. Low Temp. Phys. \textbf{134}, 1119 (2004).
\bibitem{VoutilainenPRB} J. Voutilainen, T. T. Heikkila, and N. B. Kopnin, Phys. Rev. B \textbf{72}, 054505 (2005).
\bibitem{SukumarPRB09} S. Rajauria, H. Courtois, and B. Pannetier, Phys. Rev. B \textbf{80}, 214521 (2009).
\bibitem{comsol} Comsol multiphysics is a computer software developed by Comsol, www.comsol.com.
\bibitem{Qian} L. H. Qian, Q. H. Lu, W. J. Kong, and K. Lu, Scr. Mater. \textbf{50}, 1407–1411 (2004).
\bibitem{Timofeev} A. V. Timofeev, M. Helle, M. Meschke, M. M\" ott\" onen, and J. P. Pekola, Phys. Rev. Lett. \textbf{102}, 200801 (2009).
\end{thebibliography}
\end{document}